\documentclass[aps,prb,english,showkeys,floatfix,a4,amsmath,amsfonts,amssymb]{revtex4}
\usepackage[T1]{fontenc}
\usepackage[latin1]{inputenc}
\usepackage{amsmath,amssymb,dsfont,bm}
\usepackage{braket}
\usepackage{float}
\usepackage{graphics}
\usepackage{wrapfig}
\usepackage{amssymb}
\usepackage{graphicx}
\usepackage{bbm}
\usepackage{bm}
\usepackage{graphicx}
\usepackage{color}
\usepackage{dsfont}

\definecolor{orange}{RGB}{255,127,0}
\newcommand{\iu}{\mathrm{i}}

\newcommand{\hbare}{\hbar_{\rm eff}}
\newcommand{\jcut}{j_{\rm cut}}
\newcommand{\teq}{\!=\!}

\newcommand{\ie}{\textit{i.e.}}
\newcommand{\Tr}{\operatorname{Tr}}

\begin{document}

\title{Collectivity and Periodic Orbits in a Chain of Interacting,
       Kicked Spins}

\author{Maram Akila, Daniel Waltner, Boris Gutkin, Petr Braun and Thomas Guhr}
\affiliation{Fakult\"at f\"ur Physik, Universit\"at Duisburg--Essen,
  Lotharstra\ss e 1, 47048 Duisburg, Germany}

\date{\today{}}

\begin{abstract}
  The field of quantum chaos originated in the study of spectral
  statistics for interacting many--body systems, but this heritage was
  almost forgotten when single--particle systems moved into the
  focus. In recent years new interest emerged in many--body aspects of
  quantum chaos. We study a chain of interacting, kicked spins and
  carry out a semiclassical analysis that is capable of identifying
  all kinds of genuin many--body periodic orbits. We show that
  the collective many--body periodic orbits can fully dominate the
  spectra in certain cases.
\end{abstract}



\maketitle

\section{Introduction}
\label{sec1}

Random Matrix Theory (RMT) was developed and used starting in the 50's
to study statistical aspects of nuclei and other interacting
many--body systems, see Ref.~\onlinecite{Haake,Stock,GMW1998}.  Later on, it
was realized that RMT also works for single-particle
systems~\cite{MDK1979,CVGG1980,Berry1981,BGS1984}, prompting the
celebrated Bohigas--Giannoni--Schmit (BGS). Semiclassical analysis
revealed that the classical periodic orbits (POs) are the skeleton of
the quantum spectrum \cite{Haake,Stock,cvitano,Wintgen,welge,wintgen},
also providing strong support for the BGS conjecture
\cite{Berry1985,SieberRichter,Heusler}.  It was almost forgotten that
many--body systems were the objects of interest in early quantum
chaos. Only recently, new attempts to address many--body systems in
the present context were put forward, {\it e.g.} many--body
localization~\cite{Altshuler,Basko,Znidaric} also observed in recent
experiments \cite{bloch,bloch2}, spreading in self--bound many--body
systems~\cite{Haemmerling,Freese}, a semiclassical analysis of
correlated many--particle paths in Bose--Hubbard chains \cite{EnglI} and
a trace formula connecting the energy levels to the classical
many--body orbits \cite{EnglII,Mueller}. There are also attempts to
study field theories semiclassically \cite{cvian}. As two large
parameters exist in many--body systems, the number of particles $N$
and the Hilbert space dimension determined by the inverse effective
Planck constant $\hbare^{-1}$, different semiclassical limits are
meaningful~\cite{Osipov}.

Many--body systems show collective motion, not present in
single--particle systems.  By collectivity we mean a coherent motion of
all or of large groups of particles which can be identified in the
classical phase space as well as in the quantum dynamics. Typically, a
many--body system exhibits incoherent, \textit{i.e.}  non--collective,
motion of its particles, coherent collective motion and forms of
motion in between.  Collectivity has a strong impact on the level
statistics. While incoherent particle motion leads to RMT statistics
as in the famous example of the nuclear data ensemble~\cite{guhr_nr1,
  guhr_nr2}, collective excitations often show Poisson statistics
typical for integrable systems, as {\textit{e.g.}} in
Ref.~\onlinecite{Guhr}, see Ref.~\onlinecite{GMW1998}. Due to the mixed
phase space, the BGS conjecture is not directly applicable to
many--body systems.

To illuminate the full complexity of the motion in many--body systems
and the importance of collectivity from a semiclassical viewpoint, we
consider a chain of \(N\) interacting kicked spins. We focus on the
short time regime but consider arbitrary $N$, where the collectivity
plays a significant r\^ole. Thereby, we provide a better understanding
of spin chain dynamics as this class of systems is presently in the
focus of theoretical~\cite{Braun,Gessner,Atas,Keating} and
experimental~\cite{Simon,Neill,Kim,smith} research. This presentation
is based on our recent Letter~\cite{LettAkila}.

\section{Chain of Interacting, Kicked Spins}
\label{sec2}

Consider \(N\) kicked spins with nearest neighbor interaction as in
Ref.~\onlinecite{Prosen}, described by the Hamiltonian
\begin{equation}\label{HQM}
 \hat{H}=\hat{H}_I+\hat{H}_K \sum_{T=-\infty}^\infty\delta(t-T)
\end{equation}
with the interaction part $\hat{H}_I$ and the kick part $\hat{H}_K$,
\begin{equation}\label{Hexpli}
 \hat{H}_I=\sum_{n=1}^N\frac{4J\hat{s}_{n+1}^z\hat{s}_{n}^z}{(j+1/2)^2},
 \qquad
 \hat{H}_K=\frac{2}{j+1/2}\sum_{n=1}^N{\bf{b}}\cdot\
 \hat{\mathbf{s}}_n\,,
\end{equation}
where $\hat{\bf{s}}_n=(\hat{s}_{n}^x,\hat{s}_{n}^y,\hat{s}_{n}^z)$ are
the operators for spin $n$ and quantum number $j$. Periodic boundary
conditions, \textit{i.e.}  \(\hat{s}_{N+1}^z\!=\!\hat{s}_{1}^z\), make
the system translation invariant. Moreover, \(J\) is the coupling
constant and \(\mathbf{b}\) a magnetic field, assumed without loss of
generality to have the form ${\bf{b}}=(b^x,0,b^z)$.  The kicks act at
discrete integer times $T$.  The one period time--evolution (Floquet)
operator reads
\begin{equation}
 \hat{U}=\hat{U}_I\hat{U}_K, \qquad
 \hat{U}_I={\rm e}^{-\iu (j+1/2)\hat{H}_I}, \qquad
 \hat{U}_K={\rm e}^{-\iu(j+1/2)\hat{H}_K},
\end{equation}
where \((j+1/2)^{-1}\) takes on the r\^ole of the Planck constant
$\hbare$.  We find the corresponding classical system by replacing
\(\hat{\mathbf{s}}_m\to\sqrt{j(j+1)}\,\mathbf{n}_m\) with a classical
spin unit vector \(\mathbf{n}_m\) precessing on the Bloch sphere.  The
time evolution can therefore be interpreted as the action of two
subsequent rotation matrices
\begin{equation}
{\bf n}_m(T\!+\!1)=
R_{\bf z} \big( 4J \chi_m\big)
R_{\bf b}\big(2|\mathbf{b}|\big)
{\bf n}_m(T),
\label{eq:ClassRot}
\end{equation}
first around the magnetic field axis and then around the \(z\) axis
(Ising part) with angle \(4J\chi_m\), \(\chi_m\teq n^z_{m-1}+n^z_{m+1}\).  The
classical system can be cast in Hamiltonian form,
\begin{eqnarray}\label{class}
H({\bf{q}},{\bf{p}})=\sum_{n=1}^N\left[4J p_{n+1}p_n+\sum_{T=-\infty}^\infty\delta(t-T)
    2\left(b^zp_n+b^x\sqrt{1-p_n^2}\cos q_n\right)\right]\,,
\end{eqnarray}
from which the canonical equations follow.  The $N$--component vectors
$\bf{p}$ and $\bf{q}$ are the conjugate momenta and positions of the
$N$ (classical) spins, respectively. The vectors on the Bloch sphere
are given by
\begin{equation}\label{normvec}
 {\bf n}_m=\left(\sqrt{1-p_m^2}\cos q_m,\sqrt{1-p_m^2}\sin q_m, p_m
\right)
\end{equation}
in terms of the canonical variables. In our study, the magnetic
field \(\mathbf{b}\) has a sizeable angle with the $z$ axis to
ensure non--trivial chaotic motion. 
\begin{figure}[ht]
\vspace{0.4cm}
\includegraphics[width=0.8\textwidth]{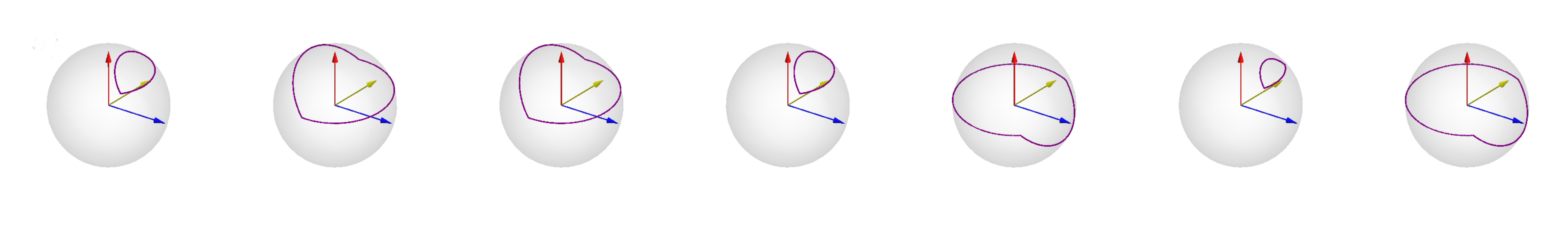}
\caption{Example for the classical motion of $N=7$ spins. Periodic orbits
         for $T=1$ kick.
         \label{fig1}}
\end{figure}
An example for the classical periodic orbits is shown in
Fig.~\ref{fig1} in the case of $N=7$ spins and $T=1$ kick.

\section{Semiclassics and Periodic Orbits}
\label{sec3}

In Ref.~\onlinecite{Waltner} we recently expressed the trace of
the propagator \(\hat{U}\) to power \(T\) for an interacting spin
system in a Gutzwiller--type--of form valid in the limit
$j\rightarrow\infty$,
\begin{equation}\label{trafor} {\rm Tr}\,\hat{U}^T\sim\sum_{\gamma(T)}
  A_\gamma{\rm e}^{\iu (j + 1/2) S_\gamma}\,.
\end{equation}
This is a sum over classical periodic orbits (POs) \(\gamma\) of
duration \(T\) if they are well isolated.  Here, $S_\gamma$ is the
classical action and, for an isolated orbit, $A_\gamma$ the stability
amplitude.  For the Hamiltonian \eqref{class}, most POs are neither
fully stable nor unstable.  The connection between the classical and
the quantum system is revealed by the Fourier transform \(\rho(S)\) of
Eq. \eqref{trafor} in $j$. This is methodically similar to
Ref.~\onlinecite{welge, wintgen} and was also used for the kicked top
by \cite{Kus,cat_schomerus}. We find
\begin{eqnarray}\label{lengthspec}
 \label{eq:sFT}
 \rho(S) =\frac{1}{j_{\rm cut}}\sum_{j=1}^{j_{\rm cut}}{\rm e}^{-\iu (j+1/2)S}{\rm Tr}\,\hat{U}^T
{\stackrel{j_{\rm cut}\to\infty}{\sim}} \,
 \frac{1}{\jcut}\sum_{\gamma(T)} A_\gamma\,\delta(S-S_\gamma)\,,
\nonumber
\end{eqnarray}
which approximates the action spectrum by peaks of width approximately
\(\pi/\jcut\) whose positions are given by the actions modulo \(2\pi\)
of the POs with length \(T\).

\section{Explosion of Dimension and Duality Relation}
\label{sec4}

\begin{figure}[ht]
\vspace{0.4cm}
\includegraphics[width=0.4\textwidth]{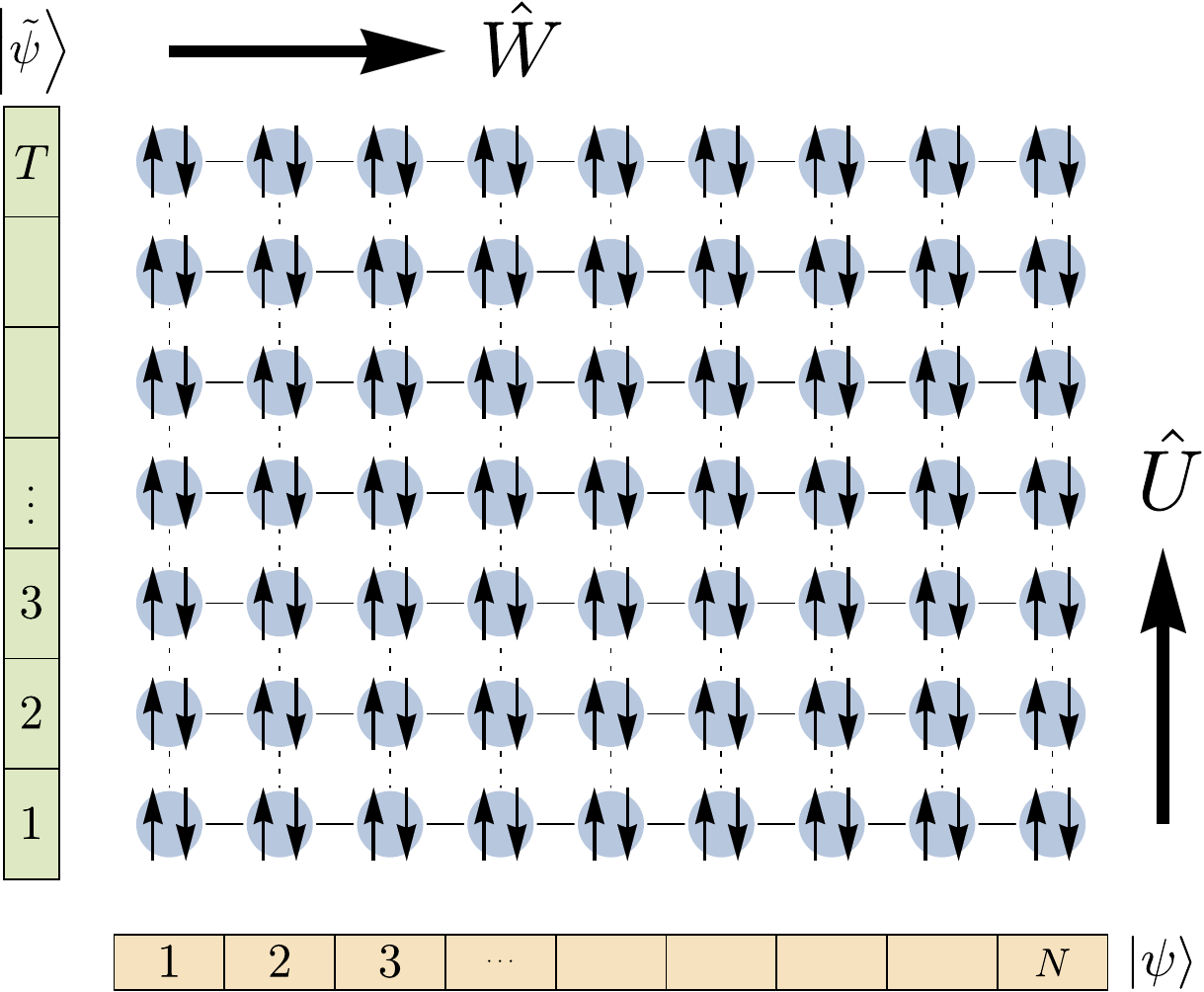}
\caption{Attempt to visualize the duality relation (\ref{dual}).
         \label{fig2}}
\end{figure}

At this point, we have to overcome a severe problem. To resolve the
peaks in \(\rho(S)\) we need to compute ${\rm Tr}\,\hat{U}^T$ for
sufficiently large \(\jcut\).  But as its matrix dimension
$(2j+1)^N\times (2j+1)^N$ grows exponentially with \(N\), a direct
calculation of the spectrum of $\hat{U}$ is impossible, \textit{e.g.},
even the propagator \(\hat{U}^T\) for \(N\teq 19\) spins at \(j=1\)
has a matrix dimension of \(10^9\!\times\! 10^9\). Luckily, recently
developed duality relations~\cite{Osipov,Akila} provide the solution
and make possible, for the first time, a semiclassical analysis of
genuine many--body orbits.  The crucial ingredient is the exact
identity
\begin{equation}\label{dual}
\Tr \hat{U}^T=\Tr \hat{W}^N\, .
\end{equation}
The trace over the time--evolution operator $\hat{U}$ for $T$ time
steps equals the trace over a nonunitary ``particle--number--evolution''
operator $\hat{W}$ for $N$ particles. Its dimension \((2j+1)^T \times
(2j+1)^T\) is governed by $T$ instead of $N$. A cartoon--type--of
visualization of the duality relation is given in Fig.~\ref{fig2}.
This duality allows us to calculate \(\rho(S)\) for arbitrary \(N\) as
long as \(T\) is sufficiently short. In Refs.~\onlinecite{LettAkila,AkilaI} we
generalize this duality approach, developed for $j=1/2$ in
Ref.~\onlinecite{Akila}, to $j\gg1$.  The dual
``particle--number--evolution'' operator is a product as well,
\(\hat{W}\teq \hat{W}_I \hat{W}_K\).  We give its explicit form using
a \((2j+1)^T\) dimensional product basis in spin space,
\begin{equation}
  |\bm \sigma\rangle = |\sigma_1\rangle \otimes|\sigma_2\rangle\otimes\dots \otimes|\sigma_T\rangle\,
\end{equation}
with discrete single spin states \(\sigma_t \in
\{-j,\,\allowbreak-j+1\,\allowbreak,\ldots\, \allowbreak+j \}\).  The
interaction part is diagonal with matrix elements
\begin{equation}
\langle{\bm \sigma}| \hat{W}_I | {\bm\sigma'}\rangle
=\delta_{\bm \sigma, \bm{ \sigma'}} \, \prod_{t=1}^T
\langle \sigma_{t}| \exp{\frac{2\iu \, \boldsymbol{b}\cdot \boldsymbol{\hat{s}}}{j+1/2}}\, | \sigma_{t+1} \rangle\,.
\end{equation}
The boundary conditions are periodic, \ie\ \(T+1\teq 1\). The kick
part, however, must have a local structure
\begin{equation}
\hat{W}_K = 
\bigotimes_{t=1}^T \hat{w}_K\,,
\quad
\langle \sigma | \hat{w}_K | \sigma' \rangle
=\exp{\frac{4\iu J \sigma \sigma'}{(j+1/2)^2}}\,.
\end{equation}
Although \(\hat{w}_K\) is related to the interaction of \(\hat{U}_I\)
it is not diagonal. In the integrable case (\(b^x\teq 0\)) the
dual operator acquires are particularly simple form which we give for
illustrative purposes,
\begin{equation}
\hat{W}_{nm}=
\exp \bigg( \iu \frac{4JT}{j+1/2}(n-j-1)(m-j-1) +  2\iu T b^z (n-j-1) \bigg)\,.
\end{equation}
The indices \(m, n\) run from 1 to \(2j+1\) and time turns, in this
case only, to a value set by the system parameters.  For further
details, see Refs.~\onlinecite{LettAkila,AkilaI}.

\section{Dominance of Collectivity in Classical Action Spectra}
\label{sec5}

\begin{figure}[ht]
\vspace{0.4cm}
\includegraphics[width=0.4\textwidth]{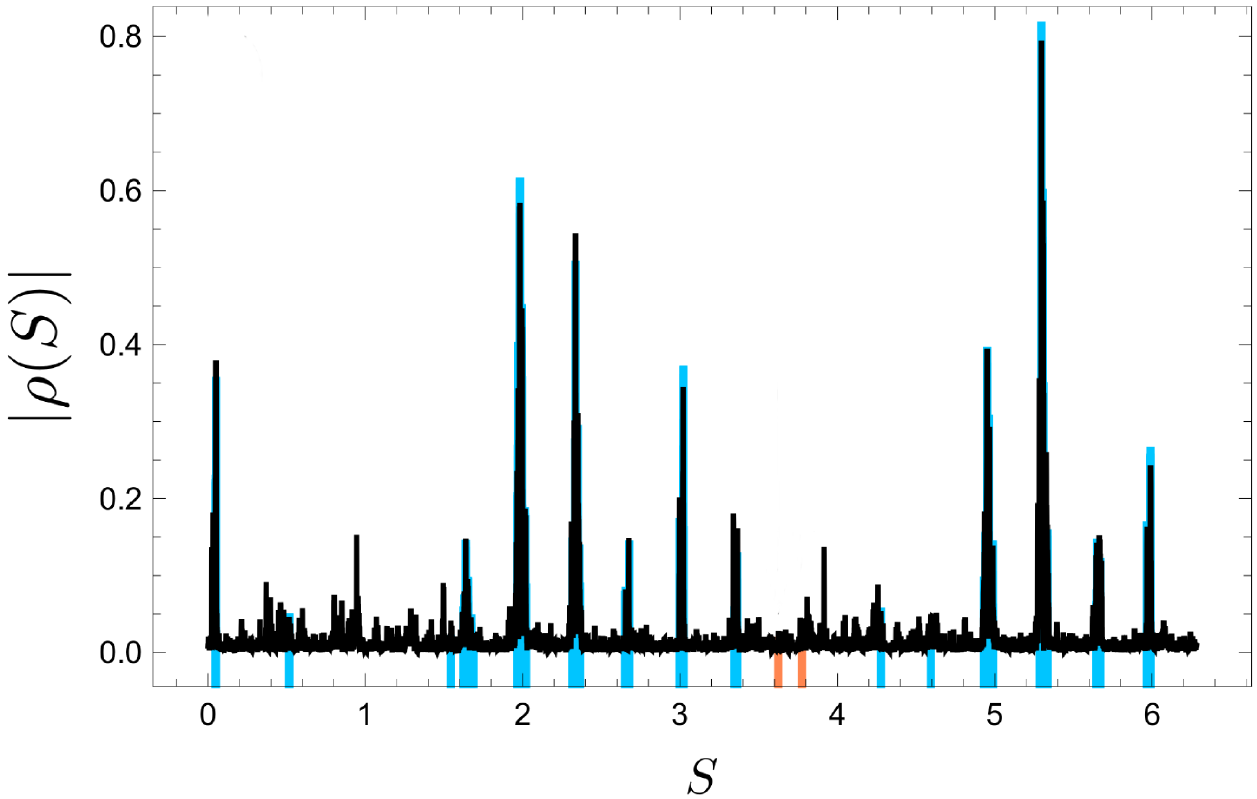}
\caption{Action spectrum for $N=19$ spins and $T=1$ kick.
         \label{fig3}}
\end{figure}

\begin{figure}[ht]
\vspace{0.4cm}
\includegraphics[width=0.4\textwidth]{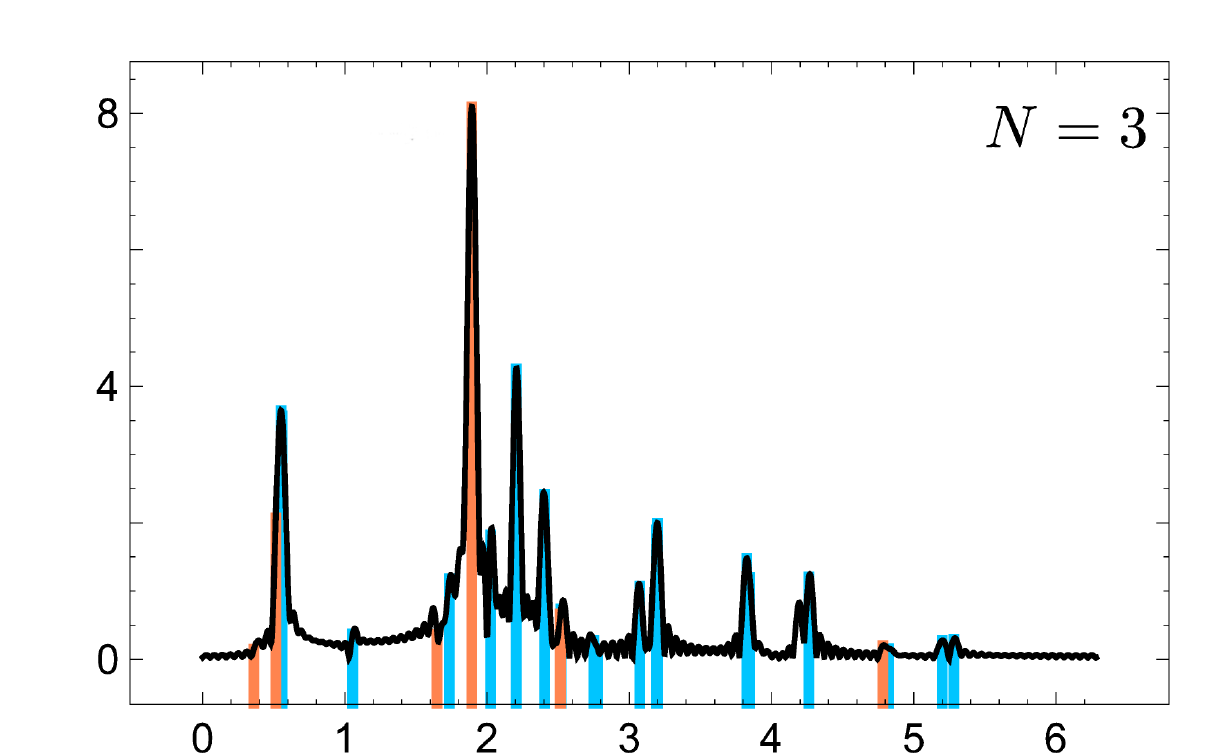}\hspace*{4mm}\includegraphics[width=0.4\textwidth]{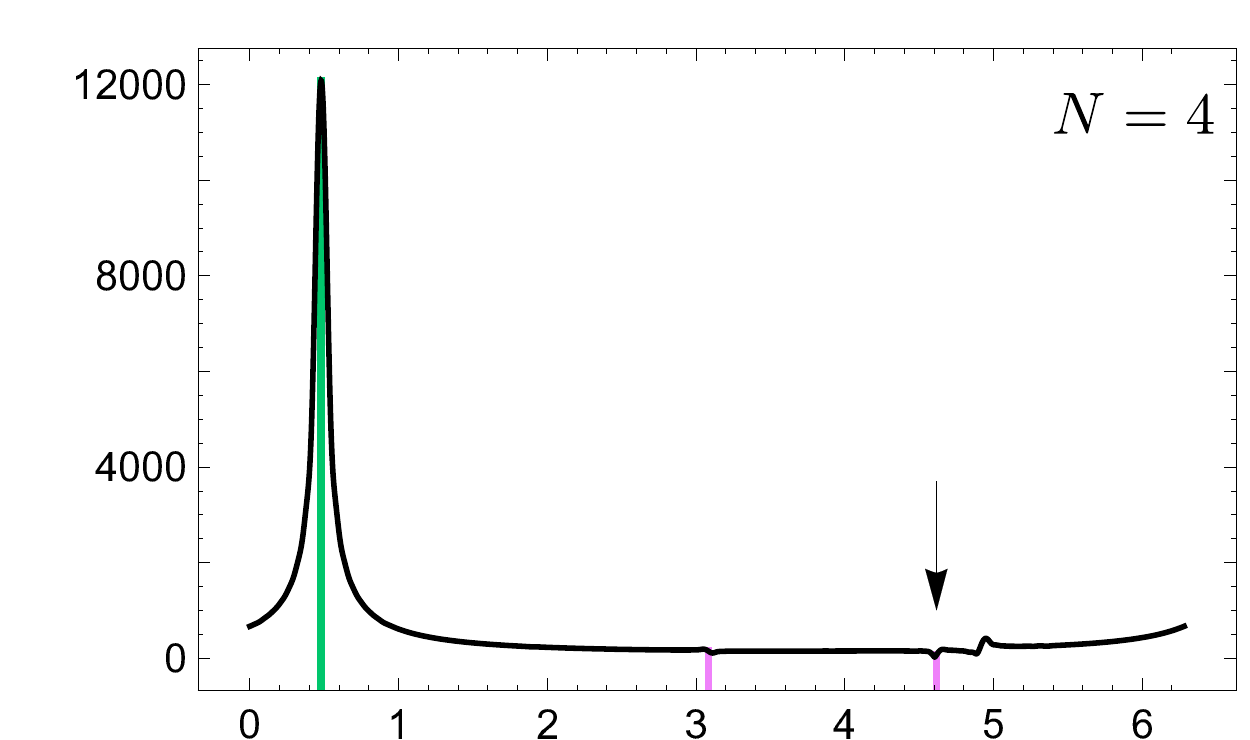}
\caption{Action spectra for $N=3$ and $N=4$ spins on the left and right hand 
         side, respectively, for $T=2$ kicks.
         \label{fig4}}
\end{figure}

\begin{figure}[ht]
\includegraphics[width=5cm]{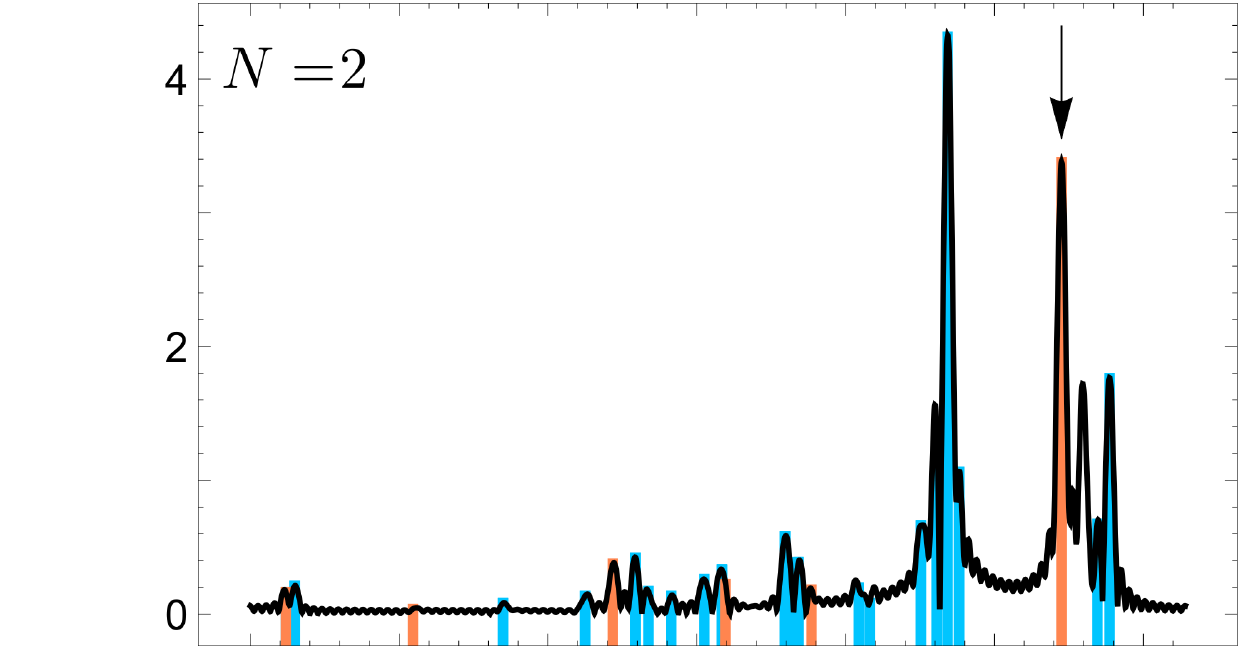}\includegraphics[width=5cm]{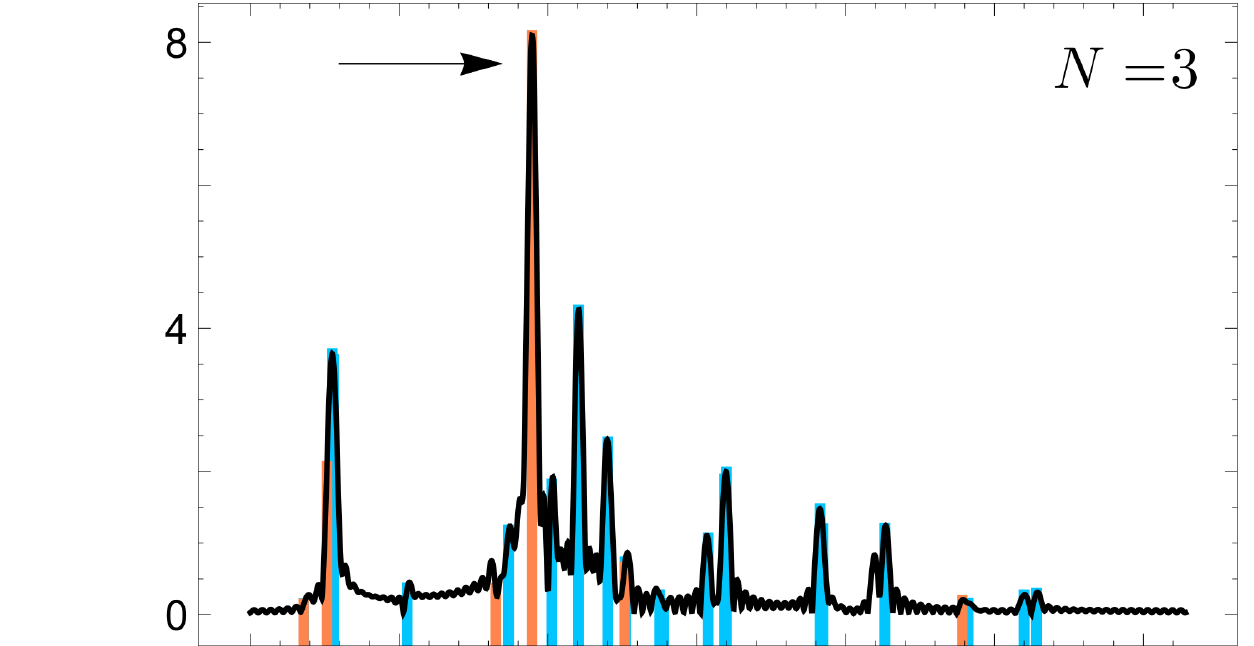}\includegraphics[width=5cm]{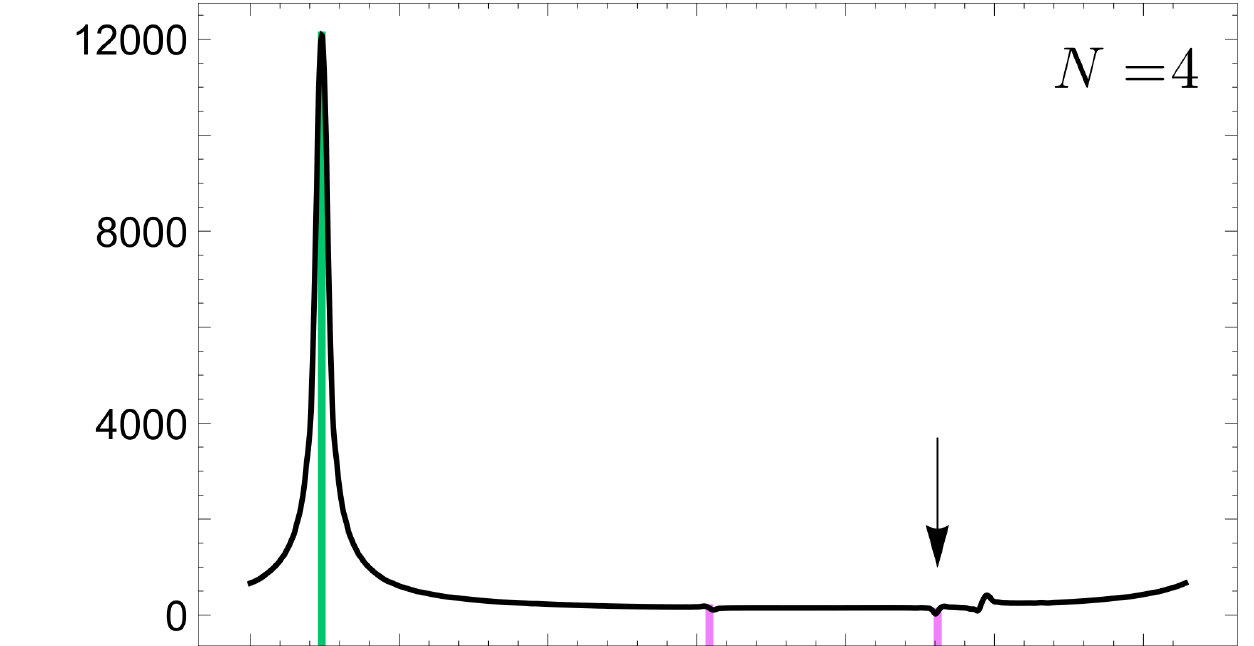}
\vspace*{-1mm}
\includegraphics[width=5cm]{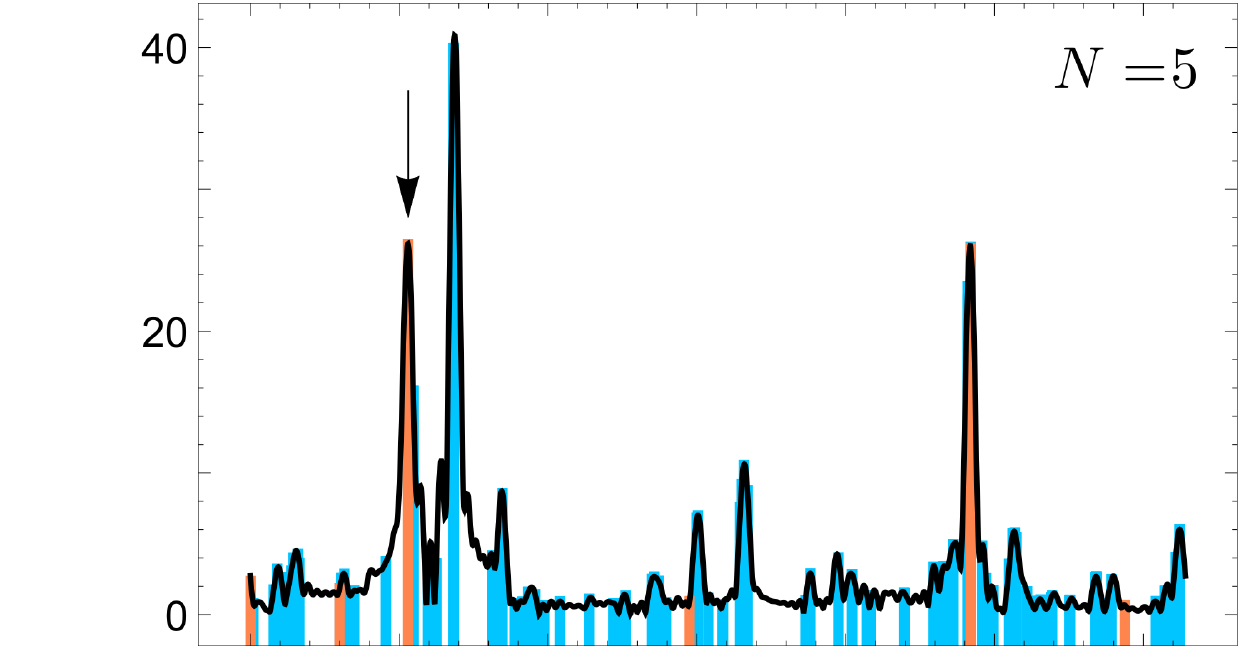}\includegraphics[width=5cm]{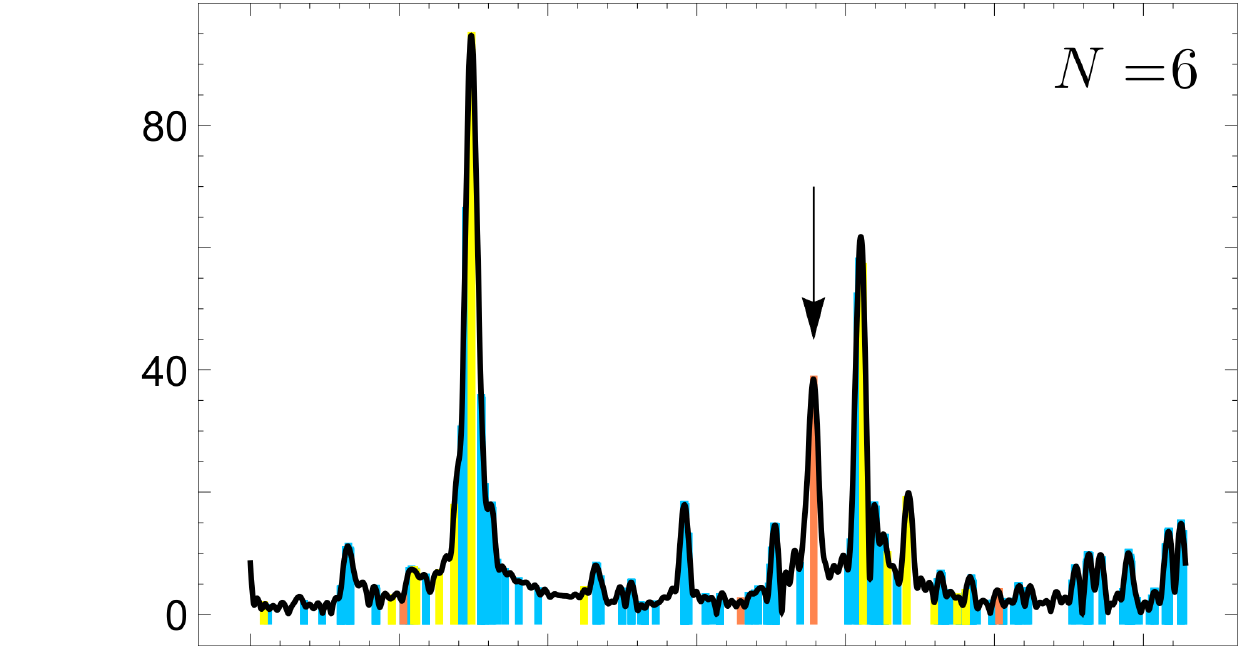}\includegraphics[width=5cm]{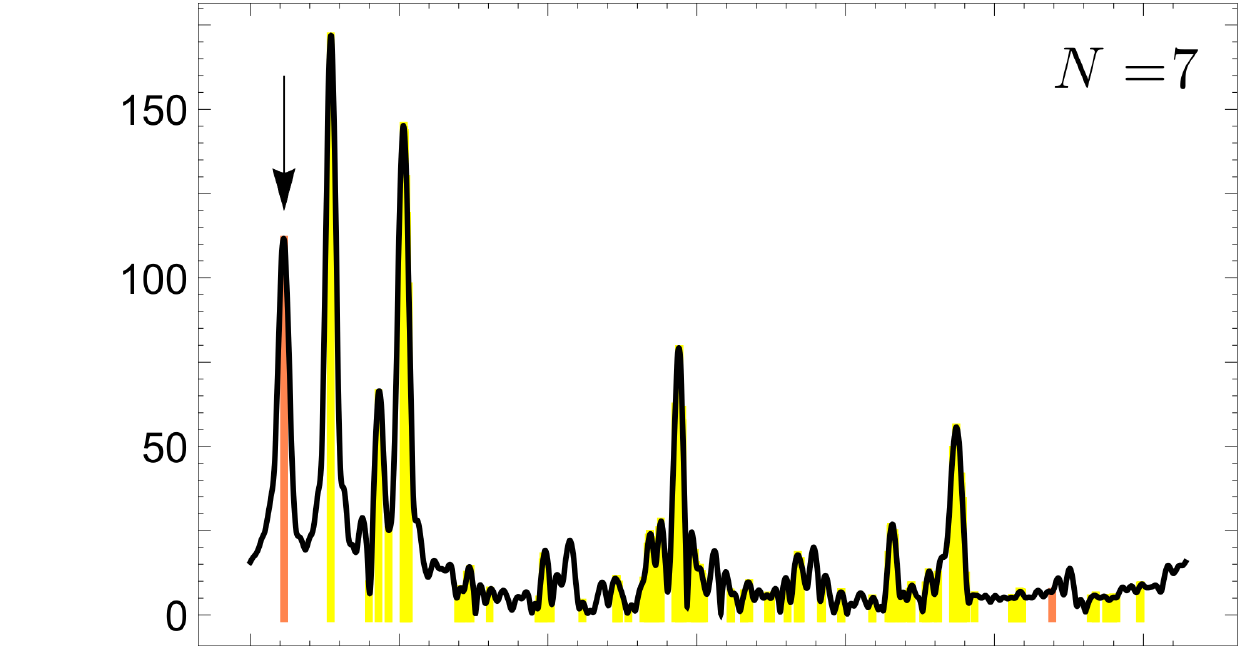}
\vspace*{-1mm}
\includegraphics[width=5cm]{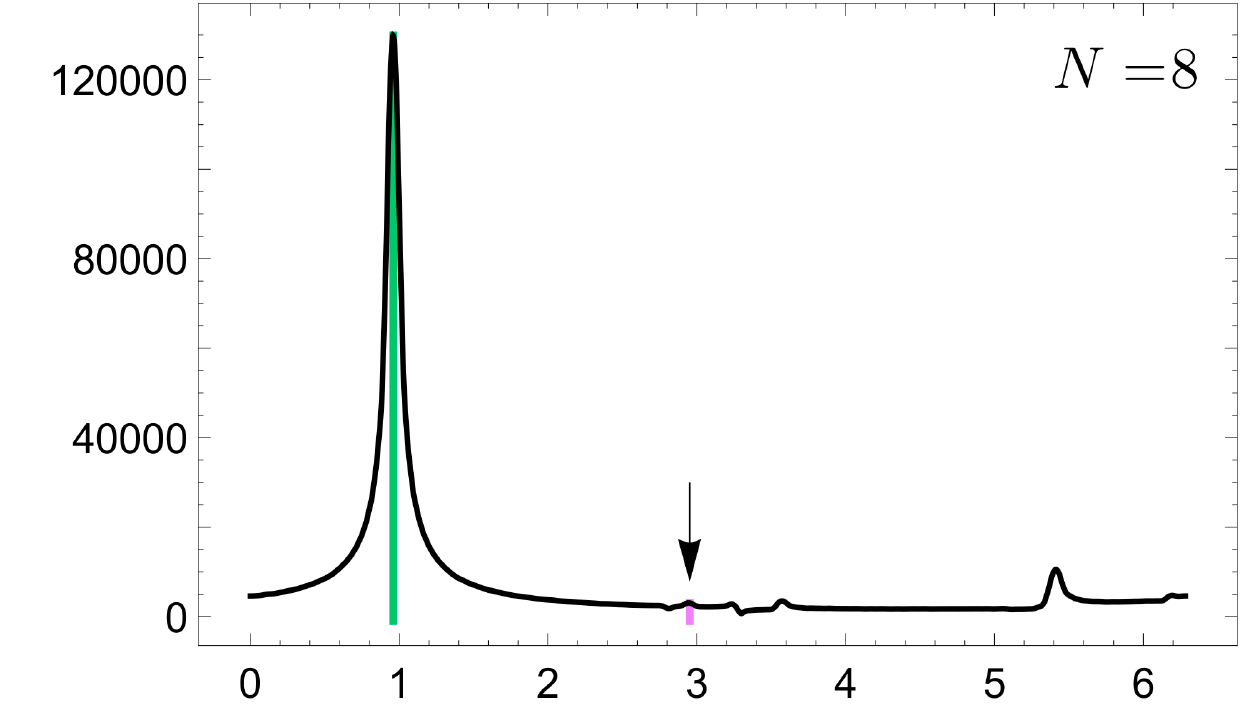}\includegraphics[width=5cm]{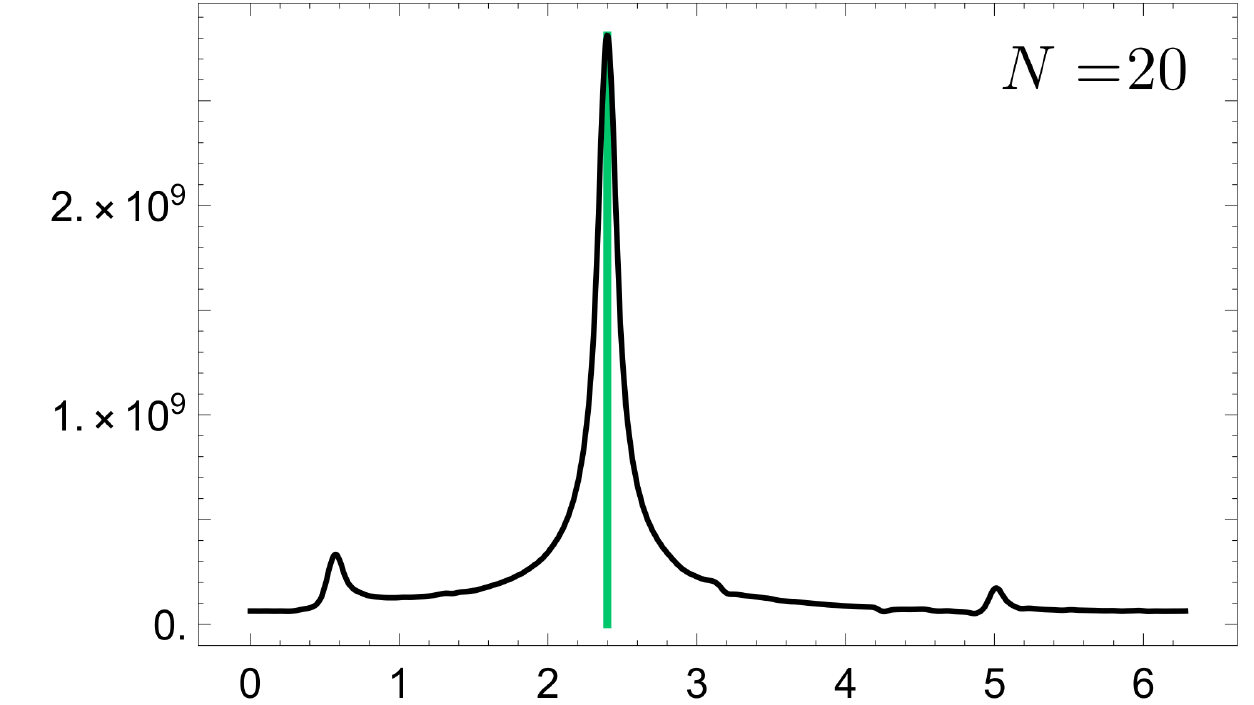}\includegraphics[width=5cm]{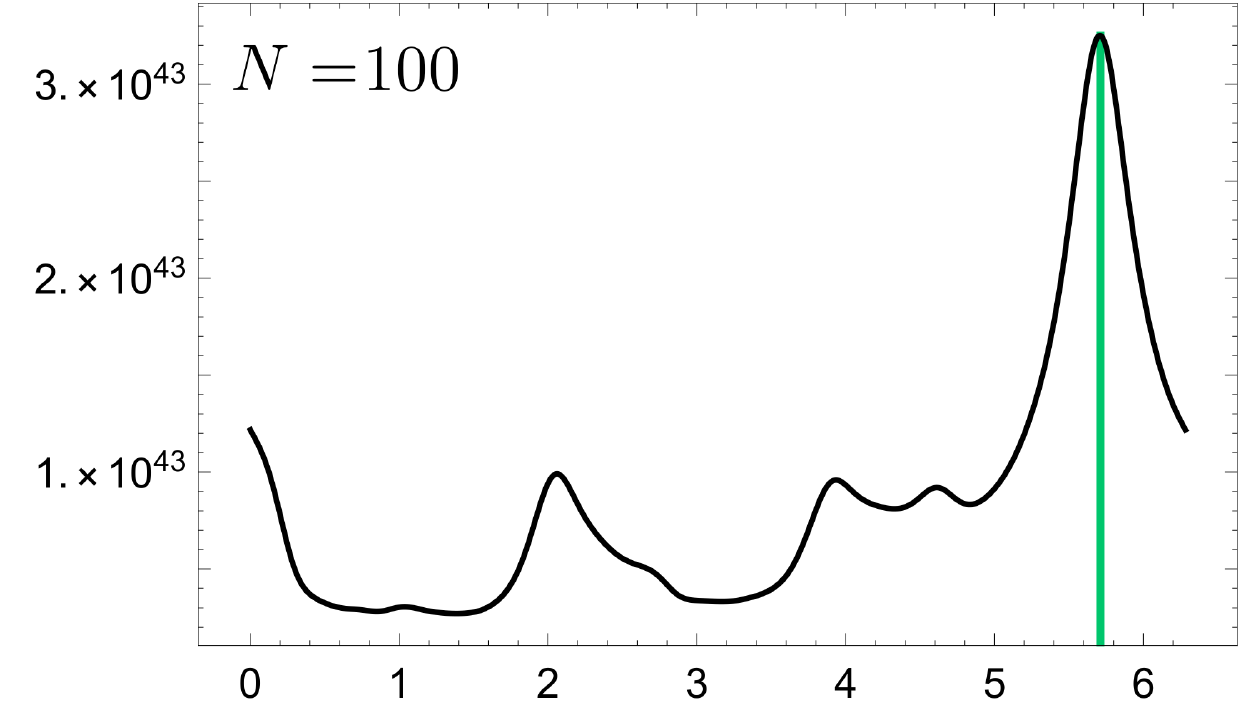}
\caption{Action spectra for $N=2,3,4,5,6,7,8,20,100$ spins for $T=2$ kicks.
         \label{fig5}}
\end{figure}

\begin{figure}[ht]
\vspace{0.4cm}
\includegraphics[width=0.4\textwidth]{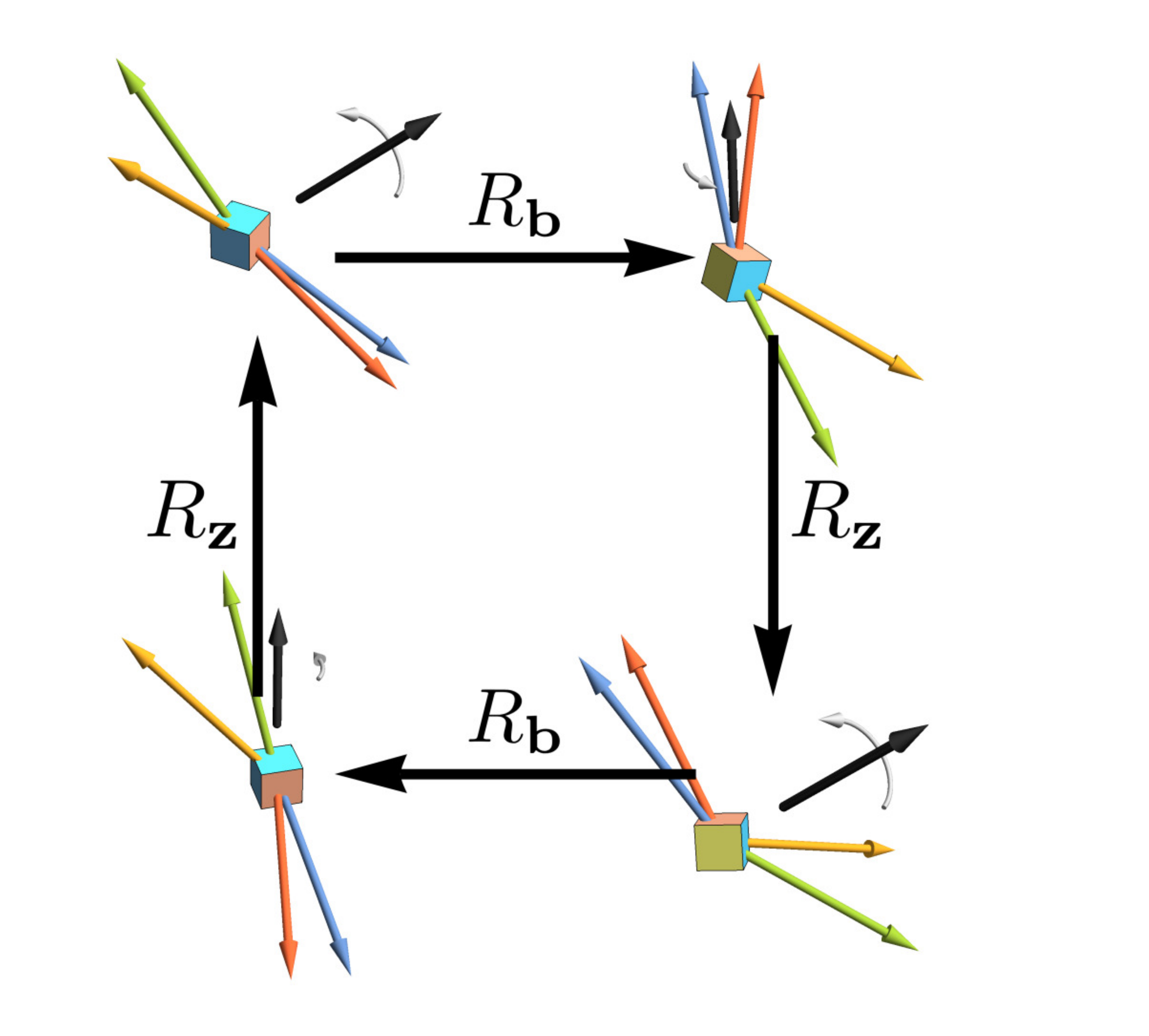}
\caption{Rigid--body--type--of rotation of all groups af four spins if
         the number of spins $N$ is an integer mutiple of four.
         \label{fig6}}
\end{figure}

We numerical calculate action spectra $|\rho(S)|$ for \(T\teq 1\) and
\(T\teq 2\) kicks, thereby exploring the short--time behavior. We
do this by, first, evaluating the traces of the
quantum mechanical time--evolution operator with the duality relation
and, second, by computing the classical periodic orbits. Hence, we
obtain the action spectra in both ways indicated in
Eq.~(\ref{eq:sFT}).  We begin with $N=19$ spins and $T=1$ kick, both
calculations are shown in Fig.~\ref{fig3}. The positions of the
periodic orbits are are indicated below the horizontal line at zero.
Very good agreement is seen, even for the peak heights. We now turn
to $T=2$ kicks. As depicted in Fig.~\ref{fig4}, the action spectra
differ strongly for $N=3$ and $N=4$ spins. We are led to argue that,
in the case $T=2$, the motion for $N=3$ spins is largely incoherent
motion of the spins, while it is coherent and collective for $N=4$
spins. This can be understood by looking, always in the case $T=2$, at
the action spectra for a varying numbers of spins in
Fig.~\ref{fig5}. Whenever the number $N$
of spins is an integer multiple of four, the spectra are dominated by
one very large peak which is much higher than in the case of the other
numbers $N$ of spins. Careful analysis of the classical phase space
yields an explanation by revealing the occurrence of four--dimensional
manifolds of non--isolated periodic orbits with equal actions. The
effect is illustrated in Fig.~\ref{fig6}. Whenever the number $N$
of spins is an integer multiple of four, the spins organize themselves
into subgroups of four spins each which perform a
rigid--body--type--of rotation in which these four spins do not
exhibit any kind of relative motion. This is a strongly coherent,
collective motion which, as Fig.~\ref{fig5} shows, outpowers the
individual incoherent motion, completely dominating the action
spectra. This phenomenon cannot be isolated. We expect similar, yet
geometrically different, forms of collective motion for other numbers
$N$ of spins and other numbers $T$ of kicks.

\section{Conclusions}
\label{sec6}

We carried out a semiclassical analysis of a (non--integrable)
interacting, many--body quantum system. We studied a kicked spin chain
representing a class of systems presently being in the focus of
experimental and theoretical research. For the first time, we
presented a unifying semiclassical approach to incoherent and to
coherent, collective dynamics. Such an interplay between different
kinds of motion is common to very many, if not all, large systems. The
key tool was a recently discovered duality relation between the
evolutions in time and particle number. It outmaneuvers the
drastically increasing complexity of the problem with growing particle
number. In the spin chain a certain type of collective motion strongly
dominates the spectra, whenever the particle number is an integer
multiple of four. An experimental verification is likely to be
feasible in view of the improving ability to control systems with
larger numbers of spins.

\section*{Acknowledgement}

One of us (TG) is grateful to the organizers of the 8th Workshop on
Quantum Chaos and Localisation Phenomena in Warsaw, Poland, May 2017,
where this research was presented.

\end{document}